\documentclass[11pt]{article}
\usepackage{moriond,epsfig}
\newcommand{\LEP}{\mbox{\scshape Lep}}
\newcommand{\GeV}{\ensuremath{\mathrm{GeV}}}
\newcommand{\GeVm}{\ensuremath{\mathrm{GeV/c{^2}}}}

\newcommand{\OPAL}{\mbox{\scshape Opal}}

\newcommand{\CDF}{\mbox{\scshape Cdf}}
\newcommand{\DZERO}{\mbox{\scshape D0}}
\newcommand{\TEVATRON}{\mbox{\scshape Tevatron}}
\newcommand{\WW}{\ensuremath{\mathrm{ W^{+} W^{-} }}}
\newcommand{\mw}{\ensuremath{\mathrm{M_W}}}
\newcommand{\W}{\ensuremath{\mathrm{W}}}
\newcommand{\MeVm}{\ensuremath{\mathrm{MeV/c{^2}}}}
\newcommand{\qqqq}{\ensuremath{\mathrm{ \qqb \qbq}}} 
\newcommand{\lnqq}{\ensuremath{\mathrm{ \len  \qqb}}}
\newcommand{\qqb}{\ensuremath{q{\bar q'}}}
\newcommand{\qbq}{\ensuremath{{\bar q}q'}}
\newcommand{\len}{\ensuremath{l \nbl}}
\newcommand{\nbl}{\ensuremath{{\overline{\nu}}_{l}}}
\newcommand{\nl}{\ensuremath{\nu_{l}}}
\newcommand{\lnln}{\ensuremath{l \nbl {\overline{l}} \nl}}
\newcommand{\Z}{\ensuremath{\mathrm{Z}}}
\newcommand{\ordalph}{\ensuremath{\mathcal{O}(\alpha)}}
\newcommand{\eg}{\mbox{\itshape e.g.}}

\newcommand{\KORALW}{\mbox{\ttfamily KORALW}}

\newcommand{\JETSET}{\mbox{\ttfamily JETSET}}
\newcommand{\SKI}{\mbox{\ttfamily SK-I}}
\newcommand{\SKII}{\mbox{\ttfamily SK-II}}
\newcommand{\ARIADNE}{\mbox{\ttfamily ARIADNE}}
\newcommand{\ARII}{\mbox{\ttfamily AR-II}}
\newcommand{\HERWIG}{\mbox{\ttfamily HERWIG}}

\newcommand{\QEDPS}{\mbox{\ttfamily QEDPS}}
\newcommand{\KeV}{\ensuremath{\mathrm{KeV}}}
\newcommand{\MeV}{\ensuremath{\mathrm{MeV}}}
\newcommand{\ufm}{\ensuremath{\mathrm{fm}}}
\newcommand{\LUBOEI}{\mbox{\ttfamily LUBOEI}}
\newcommand{\gw}{\ensuremath{\Gamma_{\mathrm{W}}}}
\newcommand{\NuTeV}{\mbox{\scshape NuTeV}}
\def\PLB{{\em Phys. Lett.}  B}

\def\EPJ{{\em Eur Phys.} J}

\def\be{\begin{equation}}
\def\ee{\end{equation}}
\def\bea{\begin{eqnarray}}
\def\eea{\end{eqnarray}}

\begin{document}
\vspace*{4cm}
\title{Measurement of the W Mass at LEP2}
\date{ 15th May, 2002 }
\author{  C.J. Parkes \footnote{e-mail: Chris.Parkes@CERN.CH}
On behalf of the \LEP\ Collaborations}

\address{Department of Physics and Astronomy, Kelvin Building, University Avenue,\\
Glasgow G12 8QQ, Scotland, U.K. }

\maketitle\abstracts{
The mass of the W boson has been measured by the \LEP\ collaborations from the data recorded during the \LEP2\ programme at $e^+$ $e^-$ centre of mass energies from 161 to 209 \GeV, giving the result : 
\begin{center} $\mw = 80.450 \pm 0.039~\GeVm$. \end{center} 
This paper discusses the measurements of the \W\ Mass from direct reconstruction of the invariant mass of the \WW\ decay products, particular emphasis is placed on the evaluation of systematic errors. Results on the direct measurement of the \W\ width are also presented.}

\section{Introduction}

The principal aim of the \LEP2\ measurement programme was the estimation of the \W\ Mass to a precision of better than 50 \MeVm\ \cite{yb}. This goal has been achieved by the combination of data from the \LEP\ experiments. The result is dominated by the systematic, rather than statistical,  error components.  

Approximately 35,000 \WW\ pair events have now been identified by the \LEP\ collaborations. The majority of the obtained statistics have been analysed \cite{latest}, and are included in the average presented here. However, the results presented here are preliminary as the collaborations are currently engaged in improving this measurement through a re-analysis of the data sample.  

\section{Analysis Technique}

\begin{figure}
\begin{tabular}{cc}

 \mbox{\epsfig{file=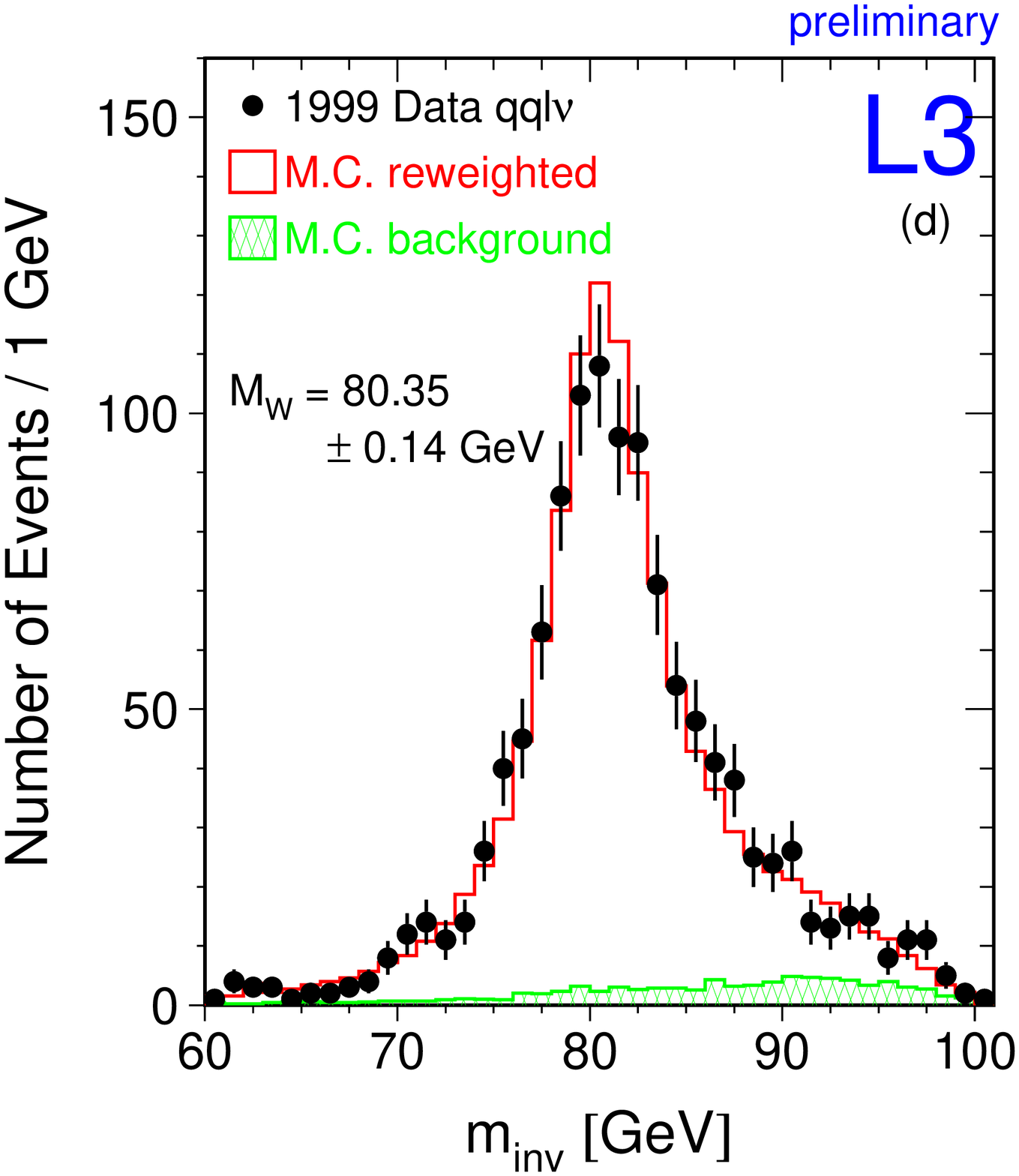,
width=0.5\textwidth,height=7cm}} &
 \mbox{\epsfig{file=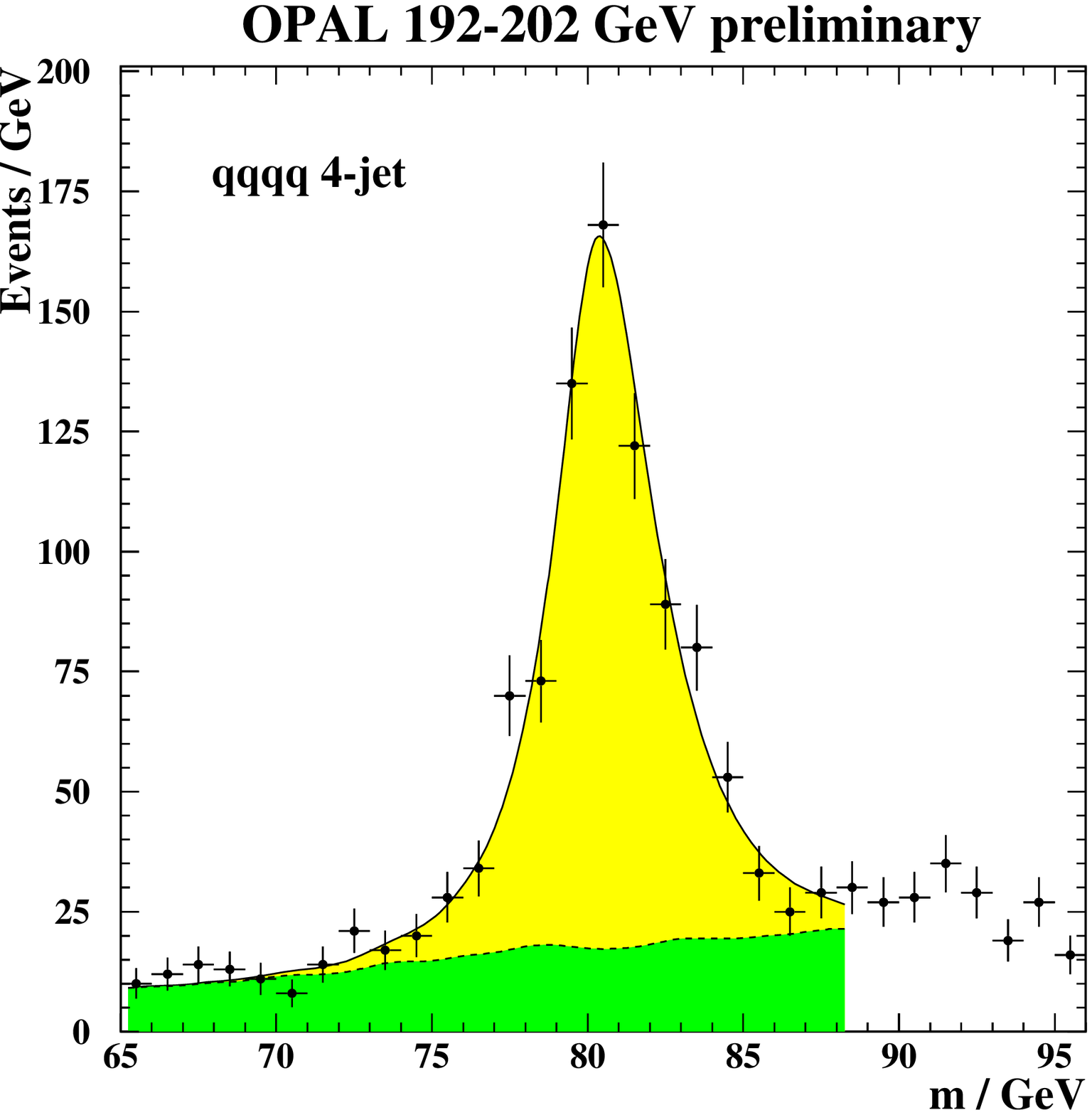,
width=0.5\textwidth,height=7cm}} \\
 (a) & (b) \\

\end{tabular}

\label{fig:mass}
\caption{Examples of reconstructed W mass distributions at \LEP\ illustrating the two main fitting techniques (see~text).}
\end{figure}

Fully hadronic (\qqqq) and semi-leptonic (\lnqq) events are primarily used for the \mw\ determination~\footnote{Fully leptonic final states \lnln\ have a relatively low mass sensitivity as a result of the presence of at least two neutrinos in the event.}. The two channels have a similar branching ratio ($46\%,44\%$)
and statistical sensitivity to \mw . The events are identified in a manner similar to that used in the \WW\ cross-section determination.

Jet clusterization is performed on the hadronic system of the event and candidate leptons from \W\ decays are identified. The \W\ boson invariant mass is then reconstructed using the observed jet and lepton four-momenta and their estimated errors.

In the \qqqq\ final state, the reconstructed jets must be appropriately paired to correspond to their parent \W s. For the case of a four-jet system three possible pairings must be considered. Five jet systems are also considered, where the fifth jet is the product of hard gluon radiation. The more sophisticated analyses make use of all possible pairings, weighting them when performing the final \mw\ fit.

The mass resolution due to detector reconstruction effects is larger than the intrinsic width of the \W\ boson. This experimental mass resolution is improved by imposing energy and momentum constraints upon the event in a constrained fit. Two highly correlated masses may be extracted for each event, or more commonly the additional constraint of equal masses is imposed.

The W Mass is evaluated by performing a maximum likelihood fit to data. Often, the fit does not rely simply upon the reconstructed event mass but rather the statistical sensitivity is improved using the chisquare of the constrained fit or the error on the reconstructed mass of the event. The probability density function used to describe the reconstructed distributions has been obtained by two techniques:  
\begin{itemize}

\item In Figure 1(a) the \lnqq\ data observed by the \L3\ Collaboration in 1999 has been compared with the distribution in simulation events. The  generated simulation events are re-weighted as a function of \mw\ using the matrix element calculation of a four-fermion generator. 

\item A semi-analytic function is used, as illustrated for the OPAL Collaboration result in Figure 1(b). This function may, for example, consist of the \W\ decay Breit-Wigner convoluted with a detector resolution function and a component describing ISR.

\end{itemize}

\section{Systematic Errors}

\begin{figure}[h]
\begin{center}
 \mbox{\epsfig{file=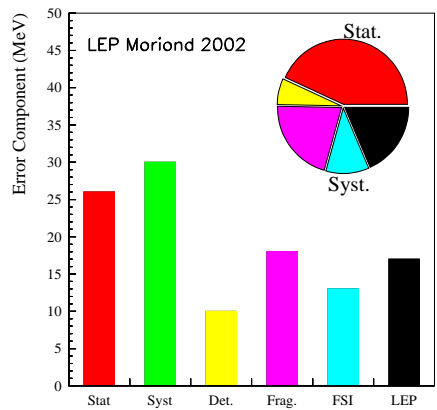,width=0.4\textwidth}}
 \label{fig:syst}
 \caption{The error on the \LEP\ \mw\ combination, the systematic error is sub-divided approximately into its main component sources of detector effects, fragmentation, \qqqq\ interconnection (FSI) and \LEP\ energy determination.} 
\end{center} 
\end{figure}

Figure 2 shows the breakdown of the systematic error on the \LEP\ W Mass combination into its major constituent elements.

\subsection{Detector Effects}

Despite the sophisticated simulation packages employed by the \LEP\ collaborations, some errors in detector modelling will remain: the result for instance of uncertainties in the tracking system alignment procedure or energy calibration of the calorimeters. These effects are most significant for the electron or muon in the \lnqq\ final state. 
Such discrepancies are studied by comparing the \Z\ peak data taken each year during calibration runs with simulation events. 
While such effects can be a significant error component for an individual experiment, they are reduced in importance in the \LEP\ combination as these error sources are uncorrelated between the experiments.

\subsection{\ordalph}

During the \LEP2\ programme \ordalph\ corrections to the four-fermion \WW\ process have become available in event generators \cite{raccoon}. These calculations include the effect of real (4f + $\gamma$ ) corrections and both factorisable and non-factorisable (\eg\ $\gamma$ exchange between decay products of different \W s) virtual corrections. These corrections have been shown to have a significant effect on the estimation of the Triple Gauge Boson couplings while the impact on the W Mass measurement is apparently much smaller. The effect of these \ordalph\ corrections will be included for the final \LEP\ W mass measurement and, while work is ongoing, it is not expected to contribute a significant additional uncertainty to the W mass estimation.

Only the modelling of the initial state real photon emission has been included in the currently assigned systematic uncertainty. At \LEP2\ energies the initial state radiation (ISR) in an event is typically several \GeV : hence the need to investigate the accuracy of the modelling of this process. The \LEP\ collaborations have studied different orders of $\alpha$ in the \KORALW\ generator treatment and performed comparisons of this treatment with an independent description (\QEDPS). All results suggest that this is not a dominant source of systematic error.

\subsection{\LEP\ Beam Energy}

The \LEP\ beam energy is applied as a constraint in the kinematic fits used in  the mass reconstruction. Hence, the fractional error on the beam energy translates directly as a fractional error on the \mw\ determination. 

At beam energies of up to 60 \GeV\ polarised electron and positron beams are produced and the beam energy may be determined extremely accurately ($\approx 200~\KeV$) through the study of the $e^+$  or $e^-$ spin precession frequency. However above this energy beam polarisation has not been achieved and the beam energy determination relies upon the accurate calibration of other methods at these low energies and their extrapolation to the \LEP2\  beam energy.
 
The current \LEP\ preliminary beam energy error is around 20 \MeV, and was assigned through a comparison of results from NMR probes and the \LEP\ flux loop \cite{guy}. The NMR probes measure the local magnetic field in several dipoles around the accelerator, while the flux loop is threaded through most of the \LEP\ dipoles and gives a measure of the integrated field. 

 This result is being cross-checked by several independent methods. A model has been developed of the accelerator's synchotron tune ($Q_s$) with RF voltage and is compared with the measurements of this process. In the autumn of 1999 the first data from the fully operational \LEP\ beam spectrometer was obtained. This project, achieved in a remarkably short time-scale, relies on the precise measurement of the bend angle of the \LEP\ beam in a specially constructed steel dipole of accurately known magnetic field. The beam energy may also be determined directly from the data using the \Z\ peak radiative return process discussed in \cite{guyradrtn}.

The production of the final energy files, used by the \LEP\ collaborations in their analyses, and the error estimate on these is expected in summer 2002. 

\subsection{Fragmentation}

The fragmentation of hadronic \W\ decays is currently the largest systematic error component. All the collaborations primarily rely for their simulation upon their own tunings of fragmentation using the \JETSET\ LUND string model. In assessing the systematic error the alternative models \HERWIG\ and the \ARIADNE\ perturbative phase model have been employed.  

The effect of varying the tuned parameters of the \JETSET\ model within their errors has also been evaluated by the \LEP\ collaborations. Each collaboration has its own \JETSET\ tuning, and work has been conducted on comparing the results of these four tunings. Uncertainties due to the baryon and kaon rates in the simulations are also being studied.

Many of the currently available results are statistically limited: to compare models to an accuracy of less than 10 \MeVm\ requires the generation of roughly one million fully simulated \WW\ events. Conservative assumptions for fragmentation have been made when obtaining the \LEP\ average.

\subsection{Final State Interactions (FSI)}

The decay distance of the \W\ bosons produced at \LEP2\ ($\approx\ 0.1~\ufm$) is significantly less than the typical hadronization scale ($\approx\ 1~\ufm$). Thus, the \qqqq\ final state may be infuenced by interactions between the final state particles arising from the two \W\ bosons.

  This is the dominant source of systematic error in the \qqqq\ channel. Although the \lnqq\ and \qqqq\ have very similar statistical sensitivity, this extra source of systematic error significantly de-weights the fully-hadronic channel in the \LEP\ combination to a $27\%$ weight. The error is assessed using the available simulation models that are compatible with the observed data.  

The developing \LEP2\ direct measurements of FSI effects \cite{fsi} will play a key role in the future estimation of this error source, certainly for Bose-Einstein Correlations and potentially also in Colour Reconnection. The current mass difference between \mw\ assessed from the \qqqq\ and \lnqq\ channels is $ +9 \pm 44 $ \MeVm, and thus gives no indication for FSI effects.

\subsubsection{Bose-Einstein Correlations}

One possible source of FSI effect is the enhanced production of identical bosons (pions) close in momentum space. The currently ascribed systematic error of  $~25~\MeVm$ is largely reliant on the \LUBOEI\ model implemented in \JETSET. This model has the correlation strength and source scale as parameters. Work on a combined \LEP\ measurement of the Bose-Einstein correlation effect is currently in progress, and indications are that the measured effect is significantly smaller than in the \LUBOEI\ model from which the W Mass systematic is currently assessed. 

\subsubsection{Colour Reconnection}

Several phenomenological models of the potentially significant non-perturbative phase reconnection effects exist. The \JETSET\ models \SKI, \SKII, \SKII ', \ARII\ and Herwig model have all been used by the collaborations. Unfortunately, current \LEP2\ measurements are not sufficiently sensitive to these effects to choose between the models. However, the current implementation of the ARII model is strongly disfavoured by \OPAL\ \LEP1\ data \cite{qqg}. Of the available models \SKI\ is generally found to give some of the largest observed shifts. The \LEP\ experiments have conservatively chosen to use this model in assessing the systematic, and choose the authors preferred reconnection parameter within this model, the assigned error is $~40~\MeVm$. In the context of the \LEP\ \WW\ workshops, the results on this model have been demonstrated to be fully correlated between the experiments. 

It is reported at this conference \cite{jorgen} that the sensitivity of the W Mass analysis to colour reconnection effects, in the \SKI\ model, can be significantly reduced through the rejection of particles in the inter-jet region. The effect on other colour reconnection models and the impact on the systematics of the \mw\ measurement remain to be studied.

\section{W Width Measurement}
The \W\ width has been measured by the \LEP\ collaborations using the same techniques, and relying on the same systematic error evaluation procedures, as for the W Mass. The \LEP\ combination gives the result

\begin{center} $\gw\ = 2.150 \pm 0.091 ~ \GeVm $.\end{center}

This can be combined with the entirely compatible direct measurement values obtained at the \TEVATRON\ by the \CDF\ Collaboration and the new \DZERO\ result \cite{D0}

\begin{center} $\gw\ = 2.23 \pm 0.17 ~ \GeVm $,\end{center}

to obtain a world average of

\begin{center} $\gw\ = 2.134 \pm 0.069 ~ \GeVm $.\end{center}

\section{Results and Prospects}

\begin{figure}
\begin{tabular}{cc}

 \mbox{\epsfig{file=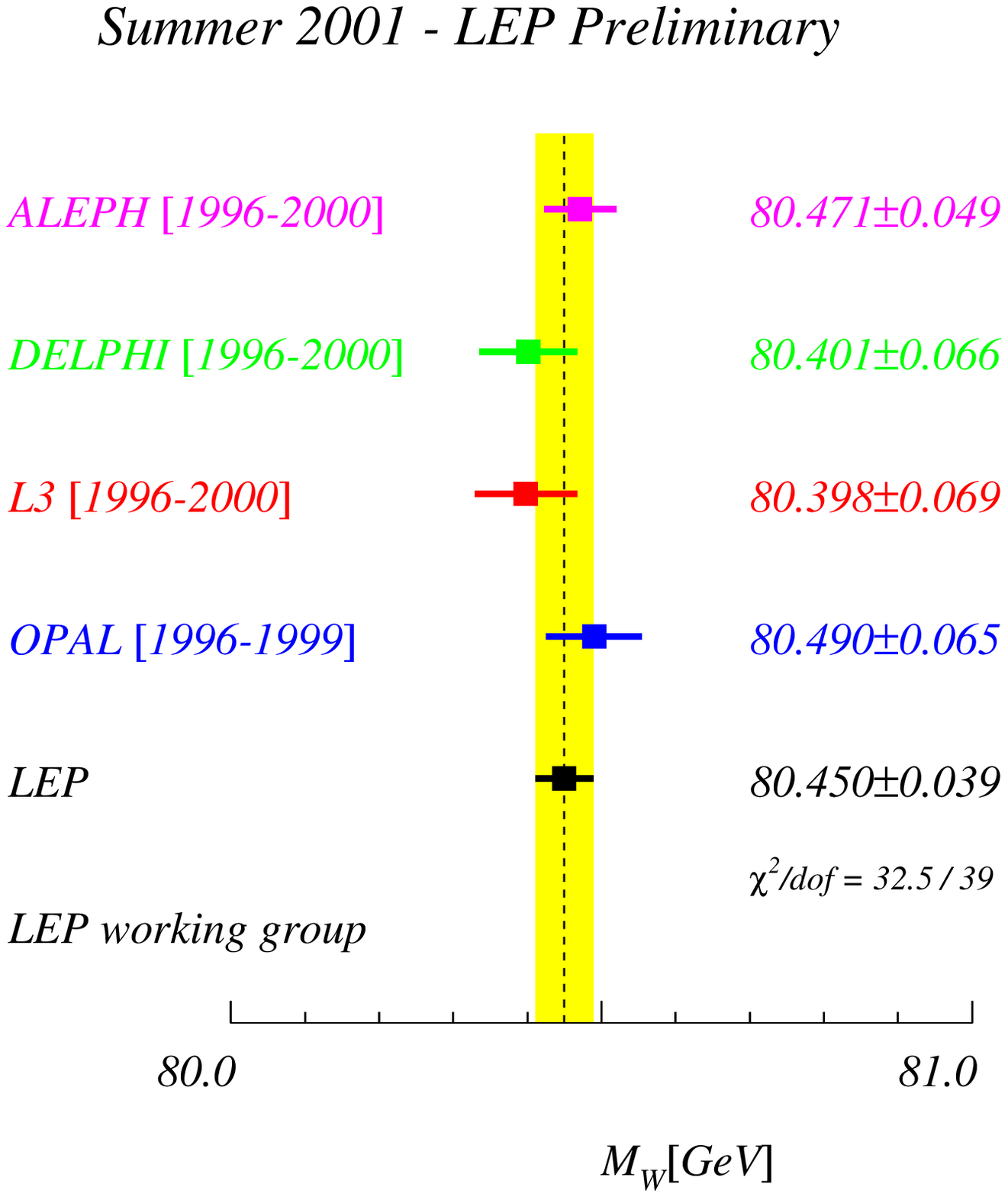,
width=0.5\textwidth,height=7cm}} &
 \mbox{\epsfig{file=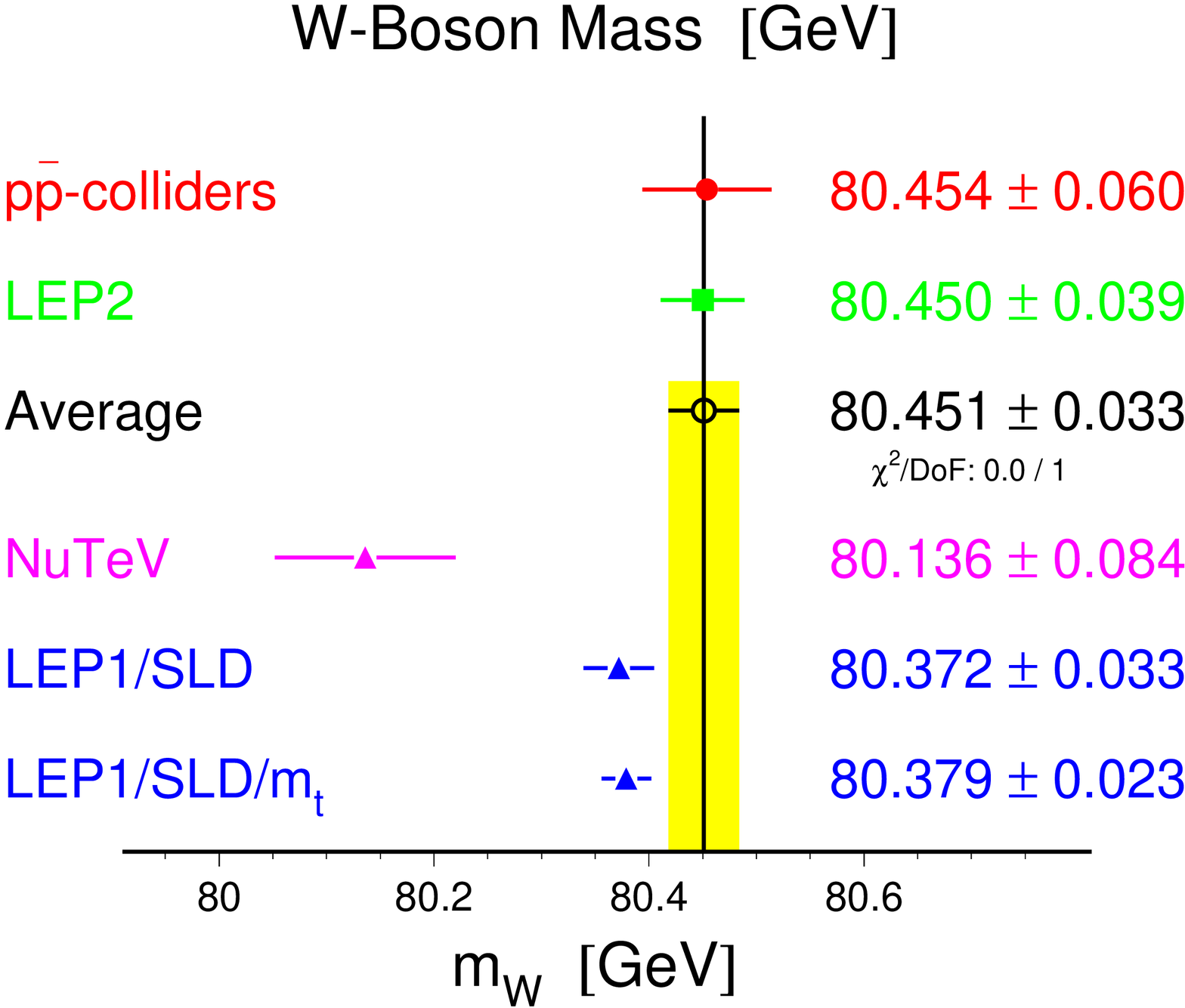,
width=0.5\textwidth,height=7cm}} \\
 (a) & (b) \\

\end{tabular}
\label{fig:mw}
\caption{The LEP2 and world average measurements of the W Mass.}
\end{figure}

The results from the \LEP\ collaborations (see figure \ref{fig:mw}) are in good agreement and the combined \LEP2\ \mw\ result in Winter 2002 is

\begin{center} $\mw\ = 80,450 \pm 26 (stat.) \pm 21 (syst.) \pm 13 (FSI) \pm 17 (LEP) ~ \MeVm$, \end{center} 

where the errors from statistical ($stat.$), final state interactions ($FSI$), LEP beam energy ($LEP$) and other systematic (syst.) sources have been quoted separately.
This value differs by 1.5 standard deviations with the indirect result obtained at \LEP1\ and 3.4 s.d. with that obtained by the \NuTeV\ Collaboration.
When combined with the highly compatible results from hadron colliders (notably the \TEVATRON ), the world average result for the direct determination is
 
\begin{center} $\mw\ = 80,451 \pm 33 ~ \MeVm$. \end{center}

The \LEP\ collaborations are currently engaged in producing the final results from the \LEP2\ data sample. These results will be based on data reprocessings incorporating improvements such as more accurate alignment constants and better tracking algorithms. Advances in the simulation of the four-fermion event generation, fragmentation tunings, and detector performance are included in the simulation events used in these analyses. The results will benefit from the higher statistical sensitivity of the fully developed analyses and our improved assessment of the systematic uncertainties.  

Hence, a modest increase in statistical precision from the current value can be anticipated. The experimental challenge will be the accurate assessment of the systematic uncertainty. In the \qqqq\ channel the FSI error
is likely to remain the limiting factor. In the \lnqq\ channel work will focus on further understanding of the fragmentation error.


\end{document}